# Induced superconductivity in the two-dimensional topological insulator phase of cadmium arsenide

Arman Rashidi, Robert Kealhofer, Alexander C. Lygo, Victor Huang, and Susanne Stemmer[*]

Materials Department, University of California, Santa Barbara, CA 93106-5050, USA.

*Corresponding author. Email: stemmer@mrl.ucsb.edu

**Abstract**

Hybrid structures between conventional, s-wave superconductors and two-dimensional topological insulators (2D TIs) are a promising route to topological superconductivity. Here, we investigate planar Josephson junctions fabricated from hybrid structures that use thin films of cadmium arsenide ($Cd_3As_2$) as the 2D TI material. Measurements of superconducting interference patterns in a perpendicular magnetic field are used to extract information about the spatial distribution of the supercurrent. We show that the interference patterns are distinctly different in junctions with and without mesa-isolation, respectively. In mesa-defined junctions, the bulk of the 2D TI appears to be almost completely shunted by supercurrent flowing along the edges, while the supercurrent is much more uniform across the junction when the $Cd_3As_2$ film extends beyond the device. We discuss the possible origins of the observed behaviors.

Interfaces between conventional, s-wave superconductors and two-dimensional (2D) topological insulators (TIs), also known as quantum spin Hall insulators (QSHIs) [1,2], can be used to induce superconductivity into the one-dimensional helical edge states of the QSHI, which is a promising route to topological superconductivity that could enable fault tolerant quantum computing [3-10]. To date, only a few experimental reports exist of induced superconductivity in QSHIs [11-15]. 2D TI-based approaches thus remain an underexplored route to topological superconductivity, despite having important advantages compared to, for example, hybrid structures with strongly spin-orbit coupled semiconductors [16].

Recently, we have shown that (001)-oriented epitaxial cadmium arsenide ($Cd_3As_2$) films with thicknesses in the range of 18 nm to 22 nm are 2D TIs [17], consistent with theoretical predictions of the effects of quantum confinement [18] on this prototype three-dimensional Dirac semimetal [18-22]. The Fermi level can be tuned by an electric gate to reside in the gap of the 2D TI [17]. The 2D electron system in these $Cd_3As_2$ heterostructures resides near the surface, which should facilitate inducing superconductivity by proximity to an s-wave superconductor that is deposited on top of the structure.

This Letter reports on investigations of superconducting hybrid structures with $Cd_3As_2$ in the 2D TI phase (previous studies of induced superconductivity in this material focused on bulk $Cd_3As_2$ in the Dirac semimetal phase [23-26]). To this end, we use Josephson junction devices, which can provide information about the induced superconductivity. For example, oscillations of the superconducting current as a function of a perpendicular magnetic field ($B$), known as superconducting quantum interference (SQI) patterns, contain information about the spatial distribution of the superconducting current density, $J_s(x)$, across the width ($W$) of the junction. For certain assumptions [27], including $\xi \gg L$, where $\xi$ is the superconducting phase coherence length and $L$ is the junction length, and a sinusoidal current-phase relation, Dynes and Fulton [28] have shown that $I_c(B)$, the maximum supercurrent (critical current) as a function of the applied magnetic field, and $J_c(x)$, the critical current distribution, are related by the modulus of the Fourier transform, i.e., $I_c(B) = \left\|\int_{-L/2}^{L/2} J_c(x) e^{ikx} dx\right\|$, where $k = 2\pi L_{eff}/\Phi_0$ and $L_{eff} = 2\lambda + L$, where $\lambda$ is the London penetration depth of the magnetic field into superconducting leads and $\Phi_0$ is the superconducting flux quantum, $h/2e$. For a supercurrent that is uniformly distributed in a rectangular junction [constant $J_c(x)$], the SQI takes the form of a Fraunhofer-like pattern, where the width of the central lobe is $2\Phi_0$, while the side lobes have $\Phi_0$ periodicity. In the case of a supercurrent that is carried only by one-dimensional edge states, such as the helical edge states of a QSHI, the SQI pattern resembles a sinusoidal double slit-like interference pattern.

The helical edge states of a QSHI can only exist at boundaries between topologically non-trivial and trivial phases, such as the physical edges of a mesa shaped device. We compare SQI patterns of junctions with and without mesa-isolation, respectively. We show that the supercurrent distribution in devices that are mesa-defined is dominated by edge modes, whereas the supercurrent flows more uniformly across the 2D bulk in devices that are not patterned.

A 20 nm $Cd_3As_2$ film was grown by molecular beam epitaxy on a $Al_{0.45}In_{0.55}Sb$ buffer layer grown on a (001) GaSb substrate, as described elsewhere [29]. A 3 nm thick GaSb layer was used to protect the surface of the $Cd_3As_2$. The topological state of the $Cd_3As_2$ film from which the Josephson junctions were fabricated was characterized using a gated Hall bar device (see refs. [17,30] for details). The gate dielectric was 25 nm of $AlO_x$ deposited by atomic layer deposition. The Landau level spectrum of the Hall bar was acquired at 1.8 K with a standard current-biased lock-in amplifier [30]. The sheet carrier density at $V_g = 0$ V was $4.8\times10^{11}$ cm$^{-2}$ and the Hall carrier mobility was 17700 cm$^2$/Vs ($V_g$ is the gate voltage). Figures 1(a) and 1(b) show the longitudinal ($\sigma_{xx}$) and Hall conductivities ($\sigma_{xy}$), respectively, obtained from tensor inversion of the resistances, as a function of a perpendicular magnetic field ($B$) and $V_g$. The filling factors at the minima of $\sigma_{xx}$ were obtained from the quantum Hall plateaus and are indicated in Fig. 1(a). At $B = 0$, a highly resistive state around -2.5 V < $V_g$ < -1.5 V appears when the Fermi level is in the gap of the 2D TI, which develops into a $\nu = 0$ plateau in $\sigma_{xy}$ at finite $B$. The crossing of two zero energy ($n = 0$) Landau levels at $B_C \sim 7$ T [Fig. 1(a)], is a hallmark of a 2D TI [31] and corresponds to the topological phase transition from inverted to conventional band order as a function of $B$. For a more detailed discussion of similar data, see ref. [17].

Fabrication of Josephson junctions involved electron beam lithography, removal of the GaSb cap layer by ion milling, and sputter deposition of 30 nm thick Ti superconducting (SC) junction leads. The junction dimensions were $L = 175$ nm and $W = 5$ μm, respectively, determined by scanning electron microscopy. Ti/Au contacts that overlapped the SC leads were deposited for wire bonding. For devices defined by mesa isolation the $Cd_3As_2$ was removed everywhere, except inside the junction and underneath the SC leads. For devices with a gate, the gate stack was similar to that of the Hall bar [17,30]. Insets in Figs. 1(c) and 1(d) show a cross-section schematic and a top-view optical microscope image of a junction, respectively. SQI patterns were measured in an Oxford Instrument Triton dilution refrigerator at a base temperature of 12 mK. Cryogenic filters were used to lower the electron temperature [32]. The differential resistance of the junctions ($R$) was obtained by adding a small AC signal (10 nA, 17.777 Hz) to the current ($I$) supplied to the junction by a Stanford Research Systems SR860 lock-in amplifier and measuring the resulting AC voltage across the junction. To measure $R$, $I$ was swept away from zero for a range of small, out-of-plane magnetic fields.

Figure 1(c) shows the superconducting transitions of the superconducting contact (Ti) and that of the induced superconductor, respectively. The transition temperatures ($T_c$) are 460 mK and 200 mK, respectively (taken when the resistance is 10% of the normal state resistance $R_N$). Figure 1(d) shows $R$ as a function of the DC bias current, which is zero in the superconducting state, spikes at the critical current ($I_C$ = 0.9 µA), before approaching the $R_N$ = 23 Ω. The induced superconducting gap, $\Delta$, is estimated from $I_C R_N \approx 20$ µeV. This yields a contact transparency, defined as $I_C R_N / \Delta_{Ti}$ where $\Delta_{Ti}$ is the gap of Ti (70 µeV), of 30%. The difference in superconducting gaps between the parent and the induced superconductors indicates that the coupling is not too strong, which could be detrimental for topological superconductivity [33].

Assuming a Fermi velocity of $v_F = 10^6$ m/s [20,34], $\xi$ is 33 µm, using $\xi = \hbar v_F / \Delta$ where $\hbar$ is the reduced Plank constant, which puts the junction into the short junction limit ($L \ll \xi$). We note, however, that $v_F$ is taken from measurements of (112)-oriented $Cd_3As_2$ and that subband or edge states in these highly confined films likely have a different $v_F$. Another indication that the junctions are in a short junction limit is the lack of hysteresis in $R$ when sweeping $I$ outward and inward, respectively. The estimated (from the Hall mobility) elastic mean free path ($l_e$) is 200 nm and the normal state phase coherence length ($l_\varphi$) is 300 nm (see ref. [35]), indicating that junctions are in the dirty limit ($l_e < l_\varphi$) and the normal state phase coherence length, $\sqrt{l_e l_\varphi}$, is 245 nm.

Figure 2(a) shows a SQI pattern of a non-mesa-isolated junction fabricated without a gate. The Fermi level is in the conduction band of the 2D TI. The observed SQI pattern is Fraunhofer-like with symmetric (with respect to $B$) and periodic lobe spacings, indicative of a relatively uniform supercurrent in the 2D bulk of the $Cd_3As_2$ thin film. Figure 2(b) shows a comparison of the data with a fit to an ideal Fraunhofer-like pattern, $I_c(B) = I(0)\text{sinc}(\pi W L_{\text{eff}} B / \Phi_0)$, which assumes a uniform supercurrent. For $W$ = 5 µm, the periodicity of the Fraunhofer pattern corresponds to $L_{eff}$ = 3 µm, which is larger than the separation of the electrodes, presumably reflecting the additional London penetration depth. The observed SQI differs quantitatively from the ideal Fraunhofer-like pattern, which predicts a reduced strength of the higher-order lobes, compared to the experiment. The discrepancy indicates the contribution of an edge mode in the SQI pattern, which also appears in the extracted spatial distribution of the critical current, shown in Fig. 2(c). The data shown in Fig. 2(c) was obtained using the Dynes and Fulton approach and the estimated $L_{eff}$. The extracted spatial current profile is slightly asymmetric, which, within the assumptions underlying the Dynes and Fulton reconstruction, is a result of the "node lifting" in the SQI pattern, i.e., the failure of the SQI nodes to reach zero. Node lifting can have many origins, topological or trivial [10,27,36], and thus there may not be a true asymmetry in the current distribution.

We next investigate a junction that was mesa isolated and had an electric gate. Figure 3 compares the ability of the gate to control the junction's channel resistance in the superconducting and normal states [Fig. 3(a)] with the gate control of the Hall bar [Fig. 3(b)]. $R_N$ of this junction is larger by about a factor of four at $V_g = 0$ compared to a junction without a gate, due to a lowering of the Fermi level after gate deposition. As already seen in Fig. 1(a), in the Hall bar geometry, $V_g$ effectively tunes the Fermi level through the gap of the 2D TI, which causes the large peak in the resistance in Fig. 3(b). In contrast, gate control is poor in the Josephson junction and $R_N$ never goes through a maximum, although the Fermi level appears to be in the gap of the 2D TI at negative $V_g$ values as $R_N$ is increasing [Fig. 3(a)]. One possible explanation for poor gate control is short-channel effects [37], which can pinch-off of the channel, resulting in the resistance saturation and destruction of induced superconductivity seen in Fig. 3(a).

Figure 4 shows SQI patterns (left column) and the extracted spatial distribution of the superconducting current (right column) at various values of $V_g$ for the mesa-isolated junction. The periodicity of the pattern corresponds to $L_{eff}$ = 2 µm, which is smaller than $L_{eff}$ of the non-mesa defined junction. If the London penetration is the same for both junctions, this would imply that the actual value of $W$ for the non-mesa-defined junction is 7 µm, instead of the nominal 5 µm, i.e., some of the supercurrent flows outside the area defined by the electrodes. The shift of the SQI patterns in Fig. 4 (central lobe is not centered at $B$ = 0) is due to a remanent flux in the magnet. This shift does not affect the calculation of the superconducting current density distribution, because it only adds a phase in the Fourier transform. We attribute the slight asymmetry of the SQI in negative vs. positive $B$ to flux trapping [38]. In addition, node-lifting is much more pronounced than for the device in Fig. 2. The node lifting can be due to non-uniformity, such as residual current paths through the 2D bulk, which leads to incomplete cancellation of the supercurrent.

The SQI pattern of the mesa-defined junction deviates from a Fraunhofer pattern. The extracted supercurrent density profiles (right column in Fig. 4) indicate edge-dominated transport with almost no 2D bulk transport. Application of increasingly negative $V_g$ does not affect the current distribution, but reduces the supercurrent density, which eventually vanishes (Fig. 3). The reduction in current density is expected, because $I_c$ scales inversely with $R_N$ [39], which increases for more negative values of $V_g$ (Fig. 3). The negligible effect of $V_g$ on the supercurrent distribution is likely due to poor gate control, discussed above, which causes the Fermi level in the channel to change very little, even at large negative $V_g$.

We next discuss the differences between the two types of junctions. In particular, the current appears almost entirely carried by edge modes in the mesa-defined junctions, while it is more uniform across the 2D bulk in case of the non-mesa junctions. While the result for the mesa defined junctions may be interpreted as suggesting the presence of helical edge states expected at the physical boundary of a QSHI, it is now known that trivial edge modes, associated with band bending effects, are common in Josephson junctions [40,41].  Further investigations are needed to determine the nature and origin of these edge modes.

Trivial origins, such as surface band bending, reflections from side surfaces, or disorder induced current paths along the sides, can be ruled out as the origin of the small edge component of the supercurrent in the non-mesa defined junction, because the $Cd_3As_2$ film extends far beyond the superconducting electrodes.  It is also unlikely that the finding is due to a misapplication of the Dynes-Fulton analysis.  Numerical calculations of SQI patterns in devices with edge-state-dominated supercurrent distributions qualitatively reproduce the results of the simpler Dynes-Fulton analysis despite relaxing the assumptions that underly it [27].  Possible explanations include the existence of intrinsic edge modes that coexist with a conducting 2D bulk or a geometrical origin.  One potential geometrical effect would be a combination of a supercurrent that extends beyond the electrodes (as discussed above) and flux-focusing.  In this case, however, an aperiodic SQI pattern should be observed [42], whereas here the periodicity is constant (Fig. 2).

To summarize, we have demonstrated hybrid Josephson junctions using 2D TI $Cd_3As_2$ thin films.  The relative contributions of edge vs. 2D bulk supercurrents in these junctions was found to be strongly determined by the position of the Fermi level and the device geometry.  All junctions showed an apparent edge mode contribution, whose nature and origins require further investigations.  The junctions may be promising for future topological quantum information devices.

The research was supported by the Office of Naval Research (Grant No. N00014-21-1-2474).  A.C.L thanks the Graduate Research Fellowship Program of the U.S. National Science

Foundation for support (Grant No. 1650114). R. K. acknowledges support through an appointment to the Intelligence Community Postdoctoral Research Fellowship Program at the University of California, Santa Barbara, administered by Oak Ridge Institute for Science and Education through an interagency agreement between the U.S. Department of Energy and the Office of the Director of National Intelligence. This work made use of the MRL Shared Experimental Facilities, which are supported by the MRSEC Program of the U.S. National Science Foundation under Award No. DMR 1720256.

**Figures and Captions**

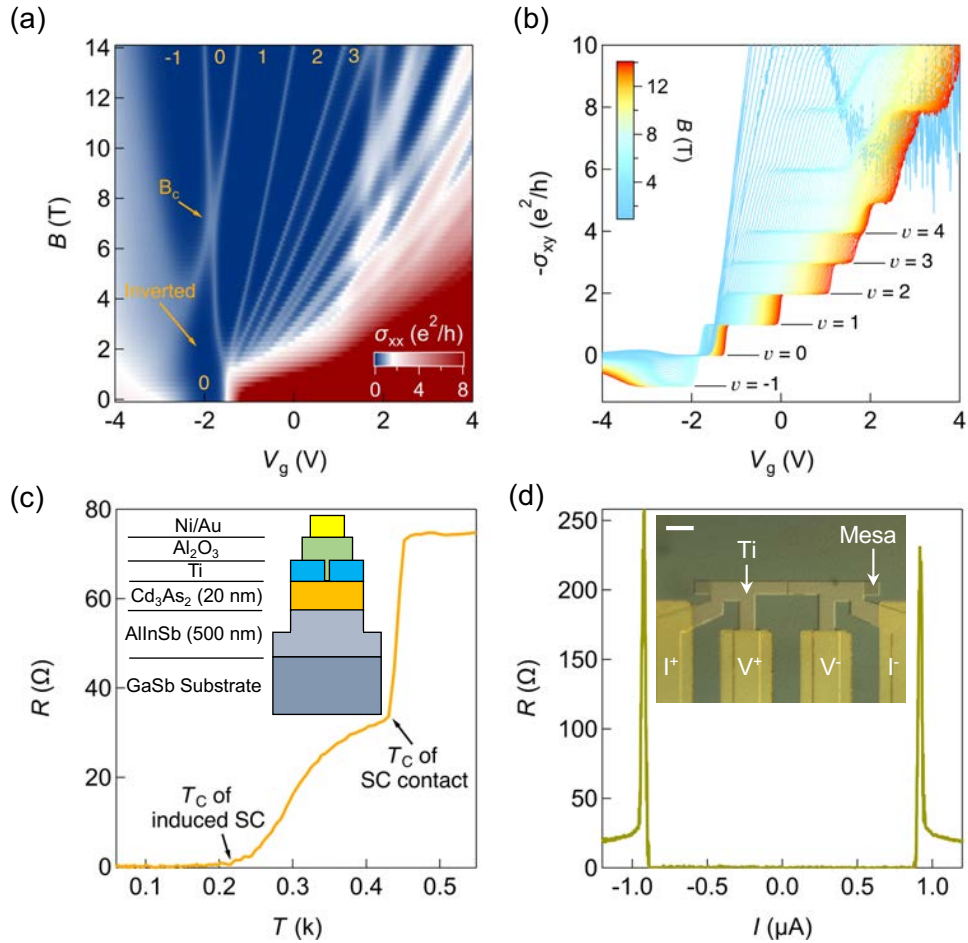

**Figure 1:** (a) Longitudinal ($\sigma_{xx}$) and (b) ($\sigma_{xy}$) Hall magneto-conductivities of the $Cd_3As_2$ film as a function of $V_g$, showing Landau levels and Hall plateaus, respectively. The labels in (a) denote the filling factors. Note the crossing of two zero energy ($n = 0$) Landau levels at $B_c \sim 7$ T. (c) Superconducting transitions. The inset shows a schematic of the heterostructure and junction. (d) Differential resistance of a Josephson junction as a function of DC current. The inset in (c) shows an optical microscope image of a Josephson junction. The scale bar is 10 μm.

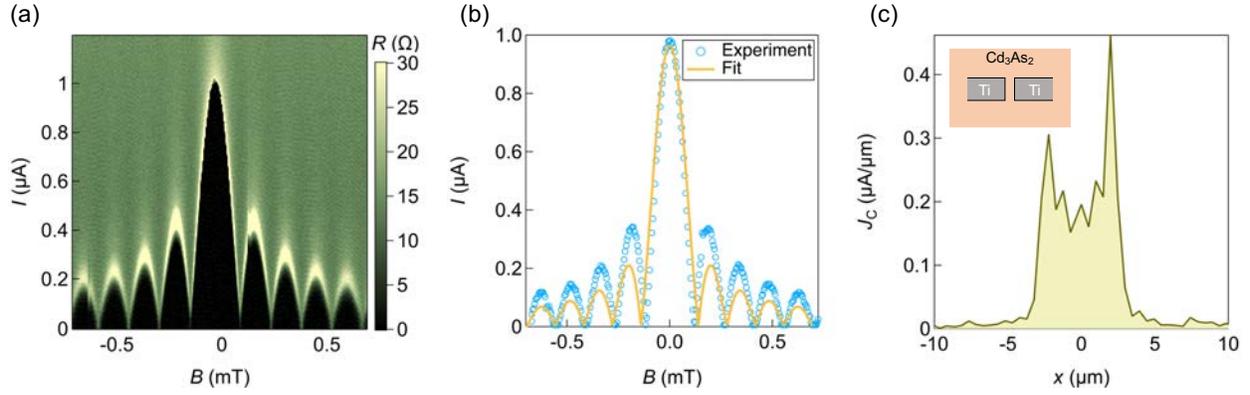

**Figure 2:** (a) SQI pattern of a Josephson junction without mesa isolation. (b) Critical current as a function of $B$ extracted from (a) and a fit to an ideal Fraunhofer-like SQI (see text). (c) Superconducting current density profile obtained using the Dynes and Fulton approach.

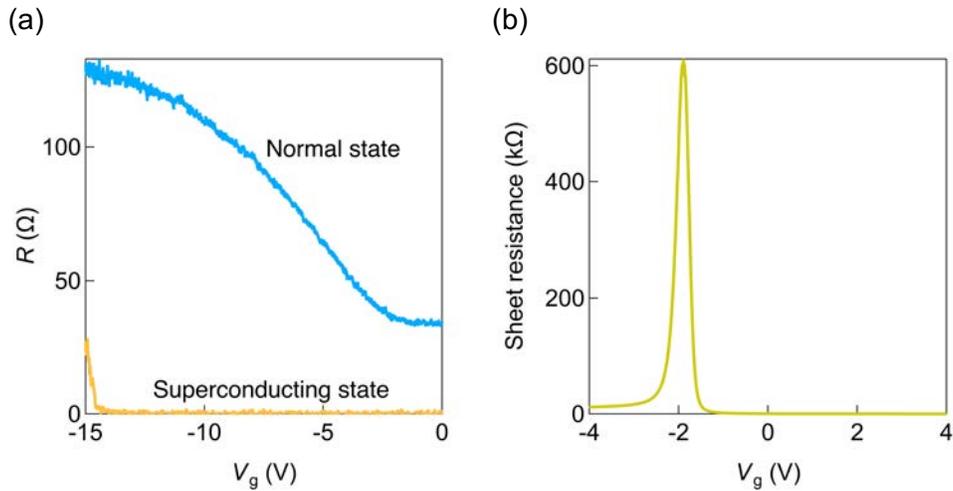

**Figure 3:** (a) Resistance of a mesa-defined, gated junction in the normal and superconducting states, respectively, and (b) longitudinal resistance of a Hall bar as a function of $V_g$.

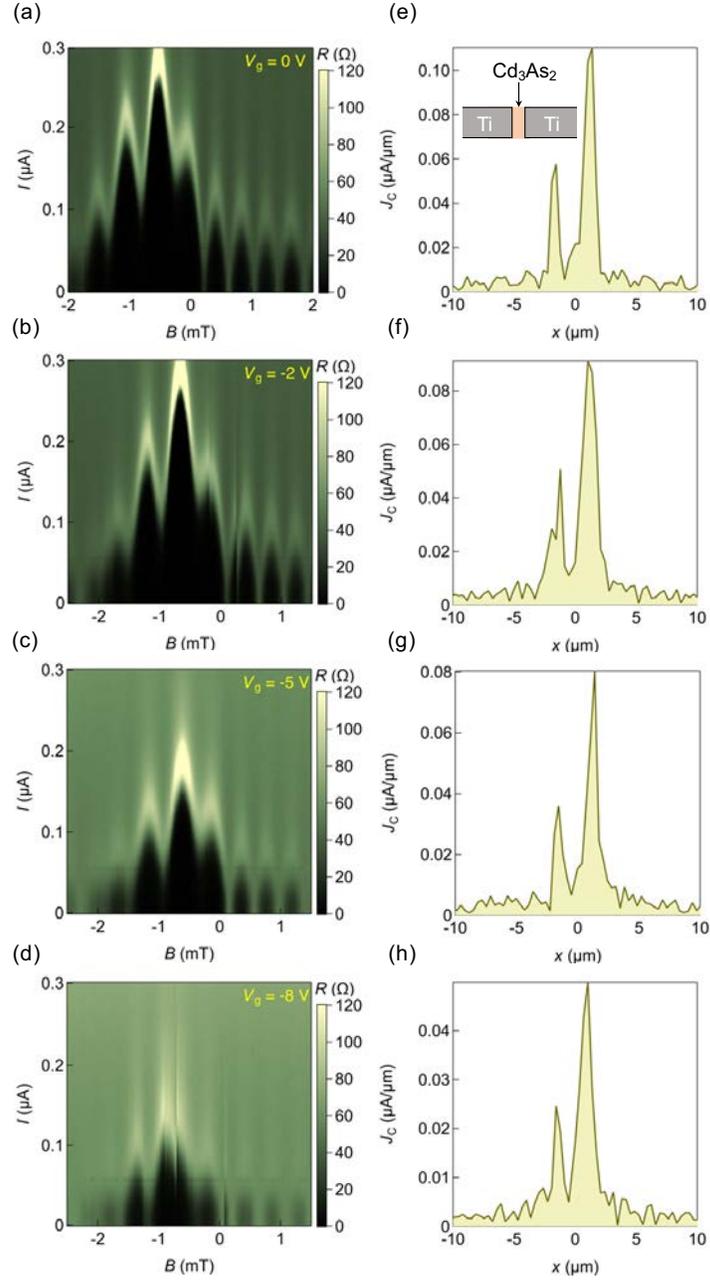

**Figure 4:** SQI patterns of a mesa-isolated junction for different values of $V_g$ (see labels) are shown on the left. The corresponding extracted superconducting current density profiles are shown on the right.